\documentclass[journal]{IEEEtran}
\usepackage{subcaption}
\usepackage{mathrsfs}
\usepackage{caption}
\usepackage{amsmath}
\usepackage{epsfig}
\usepackage{bm}
\usepackage{slashbox}
\usepackage{algorithmic}
\usepackage{latexsym}
\usepackage{amsmath}
\usepackage{amssymb}
\usepackage{epsfig}
\usepackage{amsthm}
\usepackage{graphicx}
\usepackage[ruled,vlined]{algorithm2e}
\usepackage{epstopdf}

\newtheorem{definition}{Definition}

\newtheorem{remark}{Remark}

\begin{document}
\title{\huge{Offloading in Software Defined Network at Edge with Information Asymmetry: A Contract Theoretical Approach}}
\author{
\IEEEauthorblockN{Yanru Zhang\IEEEauthorrefmark{1}, Lanchao Liu\IEEEauthorrefmark{1}, Yunan Gu\IEEEauthorrefmark{1}, Dusit Niyato\IEEEauthorrefmark{2}, Miao Pan\IEEEauthorrefmark{3}, and Zhu Han\IEEEauthorrefmark{1}}\\
\IEEEauthorblockA{\IEEEauthorrefmark{1}Department of  Electrical and Computer Engineering, University of Houston, Houston, TX 77004\\
}
\IEEEauthorblockA{\IEEEauthorrefmark{2}School of Computer Engineering, Nanyang Technological University, Singapore, 639798\\
}
\IEEEauthorblockA{\IEEEauthorrefmark{3}Department of Computer Science, Texas Southern University, Houston, TX 77004\\
\IEEEauthorrefmark{1}\{yzhang82, lliu10, ygu6, zhan2\}@uh.edu, \IEEEauthorrefmark{2}dniyato@ntu.edu.sg, \IEEEauthorrefmark{3}panm@tsu.edu\\
}
}
\maketitle
\pagenumbering{gobble}
\begin{abstract}
The proliferation of highly capable mobile devices such as smartphones and tablets has significantly increased the demand for wireless access. Software defined network (SDN) at edge is viewed as one promising technology to simplify the traffic offloading process for current wireless networks. In this paper, we investigate the incentive problem in SDN-at-edge of how to motivate a third party access points (APs) such as WiFi and smallcells to offload traffic for the central base stations (BSs). The APs will only admit the traffic from the BS under the precondition that their own traffic demand is satisfied. Under the information asymmetry that the APs know more about own traffic demands, the BS needs to distribute the payment in accordance with the APs' idle capacity to maintain a compatible incentive. First, we apply a contract-theoretic approach to model and analyze the service trading between the BS and APs. Furthermore, other two incentive mechanisms: optimal discrimination contract and linear pricing contract are introduced to serve as the comparisons of the anti adverse selection contract. Finally, the simulation results show that the contract can effectively incentivize APs' participation and offload the cellular network traffic. Furthermore, the anti adverse selection contract achieves the optimal outcome under the information asymmetry scenario.
\end{abstract}

\section{Introduction}

Nowadays, people use access various sophisticated services such as search engine, email, GPS navigation, streaming video, and online games from their mobile terminals through wireless access networks \cite{Sesia2009}. Wireless has become the primary access method for more and more people. The global mobile data traffic has reached 1.5 exabytes per month at the end of 2013, and will increase nearly 11-fold between 2013 and 2018, reaching 15.9 exabytes per month by 2018 \cite{Cisco2014}. The rapid increase in the mobile network traffic far exceeds the growth in service revenues as well as in the budgets required to address the new demands. Consequently, mobile service operators (MSOs) need to enhance their infrastructures and services in a timely and cost-effective manner to carry higher volumes of traffic and support more sophisticated services.

\begin{figure}[t]
    \centering
    \includegraphics[width=0.45\textwidth]{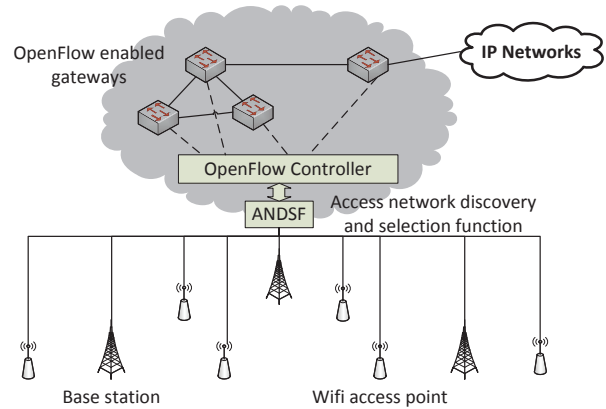}
    \caption{\label{fig:sys}An illustration of mobile data offloading enabled by the SDN-at-edge.}
\end{figure}

Mobile data offloading, which refers to moving traffic form cellular networks to alternate wireless technologies like WiFi or smallcell networks, promises to address the tremendous growth in mobile data and rapidly evolving mobile services. The mobile data offloading can be enabled by the software defined network (SDN) at edge, which is able to dynamically position or reposition the traffic in a mobile network based on various trigger criteria including the number of mobile users per base station (BS), available bandwidth, IP address and/or aggregated flow rate \cite{Manzalini2013}. The immediate advantage of SDN-at-edge is that it simplifies network management in a dense network \cite{Ali-Ahmad2013}. The software-defined radio access network (RAN) concept \cite{Gudipati2013} abstracts all independent BSs as a virtual centralized BS, which performs control plane decisions for all independent BSs at a single place. The distributed control plane in conventional RANs is therefore turned into a centralized software defined control plane \cite{Galis2013}. The system model is illustrated in Fig. \ref{fig:sys}. The central BS optimizes the network performance with global knowledge of the whole network, and can potentially adapt to the traffic variations in the network.

In this model, the access network discovery and selection function (ANDSF) can discover wireless network access points (APs) close to the mobile user and perform the mobile data offloading. The ANDSF will interact with the virtual centralized BS of the SDN for the offloading management, which can be implemented by standardized interfaces such as OpenFlow \cite{ONF2013}. Now MSOs have already deployed their own WiFi APs or initiated collaboration with existing WiFi networks to enable mobile data offloading. The SDN-at-edge can significantly alleviate both costs and operational challenges incurred by the simultaneous operation of access networks over multiple wireless technologies.

Offloading traffic through WiFi or smallcell APs is theoretically feasible. However, it is practically difficult as APs are owned by third-party and have their own traffic demand. In order to attract the APs to open their network access to cellular users, we assume that the virtual centralized BS will offer a payment based on the amount of traffic that the AP is able to offload. Meanwhile, the AP must guarantee that its own traffic is being processed first, and will only help the BS offload traffic with its idle capacity which is a private information of the APs. In practice, there exists an information asymmetry that the APs know more about their idle capacity than the BS. As we assume that all APs are selfish, they will pretend that they have large idle capacity, and thus can request more payment from the BS, which is an undesirable situation for the BS. Thus, we need to find a proper mechanism to ensure that the payments to APs match with the traffic they can offload, and thus, overcome the information asymmetry.

The model from contract theory provides us a useful tool to design such a mechanism. First, the contract can construct several traffic-payment bundles, which specify the amount of traffic that the AP needs to offload, and the corresponding payment that the BS needs to offer. Second, the contract is \emph{incentive compatible}, which includes different traffic-payment bundles that guarantees the payments to APs are in accordance with the amount of offloaded traffic. Third, the contract is \emph{self revealing}, which is designed that the APs can only achieve the maximum payoff when selecting the traffic-payment bundle that best fits into their own idle capacity. In summary, the contract theoretic model provides an incentive compatible mechanism such that the APs will select the amount of payments in accordance with the amount of traffic that they can offload for the BS, and their idle capacity will be automatically revealed to the BS as if there does not exist an information asymmetry. We name this mechanism by \emph{anti adverse selection}.

Additionally, we study another two contract mechanisms to compare with the \emph{anti adverse selection}. The first one is called \emph{perfect discrimination} in which the idle capacity of APs is available to the BS. In this scenario there does not exist the information asymmetry, the BS can treat each AP separately and offer a specific contract in accordance with its idle capacity. Thus, this mechanism can achieve the first best outcome, and serve as the benchmark of this problem. The same as the \emph{anti adverse selection} case, the second mechanism \emph{linear pricing} is also under incomplete information that the BS cannot observe the idle capacity of the APs. Unlike the \emph{anti adverse selection}, in \emph{linear pricing} the BS only specifies a unit offloading traffic per payment, and the AP chooses the payment that it wants to maximize its own payoff.

In summary, the main contributions of this paper are as follows: First we propose a novel approach to solve an incentive problem in SDN-at-edge using the framework of contract theory, which is rooted in economics research \cite{Werin.1992}. Second, we propose the \emph{anti adverse selection} to obtain an optimal results under information asymmetry, together with the \emph{perfect discrimination} and \emph{linear pricing} as comparisons. Finally, we provide thorough simulation results to prove the effectiveness of the proposed contract and show that the APs obtain incentives to offload traffic for the BS.

The remainder of this paper is organized as follows. First, we conduct a literature survey about SDN-at-edge and contract theory in Section \ref{sec:RelatedWork}. Then, in Section \ref{sec:SystemModel}, we introduce the traffic offloading trade in an SDN-at-edge system with a contract theoretic model. The problem formulation of the three contracts is well described in Section \ref{sec:ProbForm}, and we propose the solution of the three contracts. The performance evaluation is conducted in Section \ref{sec:Simulations}. Finally, Section \ref{sec:Conclusion} draws the conclusion.

\section{Related Work}\label{sec:RelatedWork}

Over the past few decades, wireless cellular network payloads have been growing fast with the introduction of smart phones, tablet computers and other new mobile devices. The development of long term evolution (LTE) is one effective way to increase the capacity of current network \cite{Sesia2009}. New technologies such as Device-to-Device (D2D) communication, WiFi, and smallcell have been introduced to offload traffic from the current wireless network. In D2D communication, devices are able to communicate with each other without requiring a dedicated wireless AP \cite{Zhang.Glob2013}. While conventional WiFi networks are typically based on the presence of controller devices known as wireless APs \cite{Camps2013}. Moreover, placing several smallcells in high dense cellular network is another widely adopted method that can offload the cellular network's traffic while be more energy efficient \cite{Quek.2013}.

As the cells tend to be smaller, and thus, the wireless infrastructure is becoming denser and heterogeneous. This gives rise to an unsustainable increase in complexity on network operations spanning across different layers due to the tight coupling in control plane decision-makings at neighboring BSs \cite{Oh2011}. It becomes necessary to design a more flexible SDN-at-edge type of architecture. While as the SDN-at-edge is controlled by a visual centralized BS, a failure of the controller can negatively compromise resilience of the whole network \cite{Valdivieso2013}. The work in \cite{Mueller2013} proposes a potential solution to find a proper traffic offloading mechanism. While this work is based on the assumption that all APs are willing to participate in the process. The novelty of our work is that we look at the problem in an economical way. Particularly, we use the framework of contract theory to model the service trading between the BS and APs.

There have been some works which try to solve the AP's incentive problem in traffic offloading that are rooted in economics. The mechanism in \cite{Gao2014} assumes that the mobile virtual network operator offers some free data quota to hosts as reimbursements (incentives) for connectivity sharing. Similarly, the work in \cite{Paris2013} investigates a bandwidth trading marketplace, where a mobile operator can lease the bandwidth made available by third parties through their APs. The work in \cite{Gao2013} also considers a market based mobile data offloading, while the system model is with multiple BSs and multiple APs. The work in \cite{Iosifidis2014} designs a distributed incentive mechanism ensures that the contribution of user, measured in the delivered mobile data, are Pareto efficient and proportionally fair, such that mobile users are willing to connect with each other and share their Internet connections. However, to our knowledge, few existing literatures adopt the contract theory to model the incentive mechanisms which can effectively attract user's participation.

Contract theory has been applied in some other areas.The work in \cite{KnapperCEC11} brings the contract theoretical model into the area of cloud computing. The authors highlight economic factors and their goal is to optimize the revenue of cloud server, taking into consideration of users' valuation of server's various characteristics in market. In \cite{Duan.INFOCOM2012}, the authors provide incentive compatible contracts to smartphone users to encourage them participating in data acquisition and distributed computing programs. In the area of cognitive radio networks, a number of works have already developed contract-theoretic techniques such as \cite{Gao.JSAC2011} and \cite{Gao.JSAC2013}. In these works, the authors model the primary user (PU) as a seller who sells spectrum resources to the secondary users (SUs) which are regarded as buyers. In \cite{Duan.DySPAN2011}, the authors proposed a system model which the PUs ``employ'' the SUs to forward their data so as to achieve higher data rates and better quality of service (QoS). However, It is difficult for the contract theoretical models in the previous works to be implemented directly into SDN-at-edge.

In summary, while traffic offloading from cellular networks have been widely studied, few literature has investigated the problem of providing incentives for APs to offload traffic in SDN-at-edge using contract theory as proposed in this paper.
\section{System Model}\label{sec:SystemModel}

Consider an SDN-at-edge with a virtual centralized BS that consists of several independent BSs, and multiple APs. Users can access the network through the independent BSs directly, or via the APs. The APs have their own traffic demand, and will only offload the cellular traffic for the BS with their idle capacity. The idle capacity of the AP is private information which is unobservable for the virtual centralized BS. The APs will trade with the central BS for helping offload the cellular traffic. We simply call the BS which refers to the virtual centralized BS in the following part. The contract that the BS offers is aiming at maximizing the offloading traffic. In the following subsections, we will first give a definition of AP type, and then model the payoffs of the BS and APs based on the contract.

\subsection{Definition of Type}

We define the AP type $\theta$ to be a representation of each AP's idle capacity which is the AP's total capacity minus the reserved capacity for its own traffic demand. Typically, high type APs can offload more traffic for the BS, and thus, are more preferred by the BS, and will receive more payment.

In practice, the idle capacity of all APs is a continuous variable. Thus, $\theta$ which represents an AP's idle capacity is a continuous value. If the BS offers every AP a specific contract, the computational complexity is high if the number of APs is large. The task will be computational complex and time consuming, thus not preferable in reality. Thus, to simplify the model from continuous case to a discrete case, we group the APs that have a similar range of idle capacity into the same type, and divide AP types into a finite number.

\begin{definition}
\label{Def1}
We divide the APs' idle capacity into $K$ types: type-1, $\ldots$, type-k, $\ldots$, type-K. We denote the types of APs by $\theta_1, \ldots, \theta_k, \ldots, \theta_K$, which are grouped in an increasing order of indices, i.e.,
\begin{equation}
\theta_1< \cdots <\theta_k< \cdots <\theta_K, \quad k \in \mathbb{Z}.
\end{equation}
\end{definition}
Here, we assume that each of the $K$ APs belongs to one of the $K$ types. Thus, the number of $AP$s in each type is $1$. For A higher $\theta$ implies a larger idle capacity to offload more cellular traffic. The BS does not know exactly the type of AP; however, it has the knowledge of the probability that an AP belongs to a certain \emph{type-k} with $k \in \{1, \ldots, K\}$ which is represented by $\beta_k$, where $\sum_{k=1}^K \beta_k=1$.

If any APs are available to users, the BS will offer contracts to those potential ``employees'' (APs) to open the network access to the users. However, different APs may have different properties (e.g. own traffic load and signal strength), the traffic offloading capacity will also differ from each other. To attract APs to offload traffic for the BS, contracts with compatible incentives must be provided. Thus, it is necessary for the BS to offer different traffic-payment bundles according to each AP's type $\theta_k$. For different APs that have different idle capacity, the BS will offer them a contract $(T(q),q)$ which includes different traffic-payment bundles. The $T(q)$ is the payment to the APs, and $q$ is the amount of traffic offloaded by the APs. For simplicity, we write the contract designed for \emph{type-k} as $(T_k,q_k)$. The APs are free to accept or decline any type of contracts. If the AP declines to receive any contract, we assume that the AP signs a contract of $(T(0),0)$. The AP chooses not to offload any traffic for the BS, and the BS will not pay the AP.

Different APs have different idle capacities, and thus, will affect their payoffs during the traffic offloading process. This in return will also affect the BS's payoff and its strategy of offering contracts in the end. In the following subsections, we will define the payoffs of the BS and APs based on the signed contract.

\subsection{Payoff of the Base Station}
In this subsection, we define the payoff of the BS when contracting with APs for traffic offloading. The BS receives benefit when traffic is offloaded via an AP, while it also has a cost on the payment to the AP. The payment to APs can be monetary or any other forms that can incentivize APs. The payoff of the BS when offloading traffic by a \emph{type-k} AP is defined as the offloaded traffic from the BS minus the cost on payment to APs, i.e.,
\begin{equation}
U(k)  = a q_k- cT_k, \quad \forall k \in \{1, \ldots, K\},
\end{equation}
where $a$ is the BS's unit monetary gain through the offloaded traffic, and $c$ is the BS's unit cost on the payment $T$. As the variable $a$ is mainly served as a transfer of the traffic gain to monetary gain to make the unit in the payoff function consistent. Thus, the numerical value of $a$ does not affect the optimization problem we will formulate. Thus, for simplicity, we assume $a=1$ here. The BS always wants to maximize its payoff by offloading the maximum traffic with the minimum payment. Apparently, the BS will not accept a negative payoff when offloading traffic through AP. In other words, we must guarantee that the trading with an AP is beneficial for the BS. Thus, we must have $q_k- cT_k \geq 0$. Otherwise, the BS will choose not to ask the AP to offload traffic.

\subsection{Payoff of the Access Point}

First, we define the valuation function $v(T)$ of the APs regarding the payments $T$, as the benefit of AP when offloading traffic for the BS. $v(T)$ is a strictly increasing concave function of $T$, where $v(0)=0$, $v'(T)>0$, and $v''(T)<0$, $\forall T$. We find that as the amount of payment increases, the satisfaction brought to the APs grows more slowly.

Then, we define the payoff of an AP of \emph{type-k} when signing a contract $(T_k,q_k)$ with the BS for the traffic offloading process as:
\begin{equation}
V(k) =  \theta_k v(T_k)-c' q_k, \quad \forall k \in \{1, \ldots, K\},
\end{equation}
where $c'$ is the AP's unit cost on offloading traffic for the BS. For simplicity, we assume $c'=1$ here. The payoff of an AP is the valuation regarding the payments minus the cost in traffic offloading. Such a cost can be power consumption, operating cost, etc.

We assume that every AP is rational. To attract the APs to participate in the traffic offloading process for the BS, the payoff that the AP receives must satisfy the following constraints.
\begin{definition}
Individual Rationality (IR): An AP will only choose to trade when the payoff that it receives is not less than its payoff that when it does not participate in the traffic offloading process, i.e.,
\begin{equation}
V(k)=\theta_k v(T_k)-q_k\geq \widetilde{V}(k), \quad \forall k \in \{1, \ldots, K\},
\end{equation}
where $\widetilde{V}(k)$ is the reservation revenue of \emph{type-k} APs when they do not take the BS's offer.
\end{definition}

Here, we use $\theta_k$ to weight the valuation of money in the utility function to represent the utility gain from the rewards. The reason for this definition is that, as we have defined $\theta$ as the AP's idle capacity, the larger idle capacity an AP has, the more interest the AP to sell those idle capacity for monetary gain. Thus, the higher type of AP with larger idle capacity, the higher valuation of money. In our system model, we normalize $\widetilde{V}=0$ without lose of generality. An AP will choose not to trade if its payoff is negative. In other words, the contract is feasible if and only if AP's payoff is equal to or greater than $0$. Clearly, the BS wants to offer the APs as little payment as possible, and leave the APs with zero payoff. However, as the BS has limited information about the APs' types, it is hard to set the contract bundle $(T_k,q_k)$ to make the AP's payoff $V(k)=0$ exactly. Usually, the BS will offer a traffic-payment bundle that brings positive payoff to the AP, which is called the \emph{information rent}.

We assume that each AP is selfish and wants to maximize its own payoff. In particular, the APs want to achieve as much payment as they can, but offload as little traffic as they could. To guarantee that the contract is incentive compatible and each AP will receive the amount of reward in accordance with the traffic they offloaded, the following constraints must be satisfied.
\begin{definition}
Incentive Compatibility (IC): Each AP can only receive the maximum payoff when selecting the contract designed for itself, i.e. type $\theta_k$ AP prefers to choose the contract $(T_k,q_k)$ than any other traffic-payment bundles.
\begin{equation}
\theta_k v(T_k)-q_k\geq \theta_k v(T_l)-q_l, \quad \forall k,l \in \{1, \ldots, K\}, \quad k \neq l.
\end{equation}
\end{definition}

Under information asymmetry, a feasible contract must satisfy the IR and IC constraints to ensure every type of APs are fully motivated, which is called \emph{incentive compatibility}. Otherwise, APs will lose the incentive to offload traffic for the BS.
\subsection{Social Welfare}

The social welfare of the network is defined as the summation of the BS and all APs' payoffs, i.e.,

\begin{align}
\Pi &=  \sum_{k=1}^K U(k)+\sum_{k=1}^K V(k), \\ \nonumber
    &= \sum_{k=1}^K [\theta_k v(T_k)- cT_k].
\end{align}
The social welfare is the difference between the benefit of traffic offloading $\theta_k v(T_k)$ and the payment cost $cT_k$. The offloaded traffic is the internal transfer between the BS and AP and is not counted in the social welfare.

\section{Proposed Solution}\label{sec:ProbForm}

In this section, we are aiming at obtaining the solution of the \emph{anti adverse selection}, which is the optimal incentive mechanism under the information asymmetry. Before that, we will present two other mechanisms as the comparisons. First, we solve the first best \emph{perfect discrimination} by considering the ideal scenario where there is no information asymmetry. Then, we will discuss about the traditional \emph{linear pricing} when information asymmetry is introduced. Finally, we will provide solution of the \emph{anti adverse selection} which brings the first best outcome under information asymmetry, and is the second best outcome when comparing with the ideal case.

\subsection{Perfect Discrimination Contract}

The \emph{perfect discrimination} contract deals with the problem of service trading between the BS and APs without information asymmetry. In particular, the BS is perfectly informed about the APs' idle capacities, i.e., the amount of traffic that they can offload from the cellular network. The BS's offloading traffic maximization is
\begin{eqnarray}\label{eq:FstBst1}
&&\!\!\!\!\!\!\!\!\!\!\! \max_{(T,q)} \sum_{k=1}^K \beta_k\left(q_k- cT_k \right),
\\
&&\!\!\!\!\!\!\!\!\!\!\!s.t. \nonumber
\\
&& (a) \quad\!\!\theta_k v(T_k)-q_k \geq 0.
\end{eqnarray}
The problem is under the IR constraint (a) that each AP obtains a payoff equal to or greater than zero which is the reservation payoff when not taking the BS's offer.  The IC constraint is unnecessary here as the BS is aware of the APs' types, and thus, the APs cannot mimic any other type APs.

As the BS can observe each AP's type $\theta$, it can treat each AP separately to solve $K$ optimization problems, defined as follows:
\begin{eqnarray}\label{eq:FstBst2}
&&\!\!\!\!\!\!\!\!\!\!\! \max_{(T_k,q_k)} \beta_k\left(q_k- cT_k \right),
\\
&&\!\!\!\!\!\!\!\!\!\!\!s.t. \nonumber
\\
&& (a) \quad\!\!\theta_k v(T_k)-q_k \geq 0, \nonumber
\\
&& \quad k  \in \{1, \ldots, K\}. \nonumber
\end{eqnarray}

For the selfish BS, it will try to extract as much payoff from the APs as they can, resulting in a zero payoff of each AP, i.e.,
\begin{equation}
\theta_k v(T_k)=q_k, \quad  \forall k\in \{1, \ldots, K\}.
\end{equation}
Replacing $q_k$ with $\theta_k v(T_k)$ and taking the first derivative of the objective function we have
\begin{equation}
\theta_k v'(T_k)=c,  \quad \forall k  \in \{1, \ldots, K\}.
\end{equation}
As $\theta_k$ and $c$ are initial values, each traffic-payment bundle $(T_k,q_k)$ can be solved by equations (10) and (11). Thus, the \emph{perfect discrimination} contract works as, first, let the BS set the payment to the AP such that the marginal valuation equals the marginal cost; second, set the amount of offloaded traffic so as to appropriate the AP's payoff as zero and leave no \emph{information rent} for the AP.

Having the optimal contract $(T,q)$, the payoff of the BS is thus
\begin{equation}
\overline{U}=\sum_{k=1}^K \beta_k\left(q_k- cT_k \right).
\end{equation}
The social welfare obtained by (6) is
\begin{align}
\overline{\Pi} =\sum_{k=1}^K \beta_k\left(q_k- cT_k \right).
\end{align}
The social welfare has the same value as the payoff of the BS. As all APs receive zero payoff during the traffic offloading process, when there is no information asymmetry. The \emph{perfect discrimination} gives the first best solution to maximize the BS's payoff. The perfect price given in this scenario is also called price of a \emph{perfect competitive market} in which the price is determined by a centralized controller \cite{Gao2013}. In this case, the social welfare and the BS's payoff achieve the Pareto efficiency and are maximized.

\subsection{Linear Pricing Contract}

Different from that in the \emph{perfect discrimination}, the BS is unobservable of the APs' types. Thus, the optimal traffic-payment contract $(T,q)$ is no longer feasible. Instead, the BS will only specify a unit of traffic per payment $P$ for the offloading process, and the APs will request the amount of payment $T$ that they want to maximize their own payoffs. The payoff of each AP becomes
\begin{equation}
\underline{V}(k) =  \theta_k v(T_k)-PT_k, \quad \forall k\in \{1, \ldots, K\}.
\end{equation}
Thus, we see that the amount of offloaded traffic $q_k=PT_k$. With a fixed traffic unit $P$, requesting more payment $T_k$ means more traffic $q_k$ to offload and more cost. Take the first derivative of the objective function regarding $T_k$, we have
\begin{equation}
\theta_k v'(T_k)=P, \quad \forall k\in \{1, \ldots, K\}.
\end{equation}
Thus, we can represent the inverse payment request function $T_k$ as $T_k=d(P,\theta_k)$. As $\theta_k$ is known, we can simply write $T_k=d_k(P)$ as a function of traffic unit $P$. The requested payment curve is a strictly decreasing function of $P$. As the traffic unit $P$ per payment is increasing, in order to lower the AP's energy/management cost $q_k=Pd_k(P)$, less payment is requested from the BS.

As the BS knows the probability $\beta_k$ that each AP belongs to a specific type $\theta_k$, it is easy to have the expected payment $D(P)$ requested by the APs, defined as follows:
\begin{equation}
D(P) =  \sum_{k=1}^K \beta_k d_k(P).
\end{equation}
Thus, the objective function of the BS changes to
\begin{align}
\underline{U}(P) &=\sum_{k=1}^K \beta_k(q_k-cT_k),  \\ \nonumber
    &=\sum_{k=1}^K \beta_k[Pd_k(P)-cd_k(P)],  \\ \nonumber
    &=PD(P)-cD(P), \\ \nonumber
    &=(P-c)D(P),
\end{align}
where $P>c$ must be satisfied such that the BS will receive a positive payoff. Otherwise, the BS will manage the traffic itself. Taking the first condition of the objective function with respect to $P$, we obtain the optimal unit traffic per payment given by
\begin{equation}
P_m=c-\frac{D(P)}{D'(P)}.
\end{equation}
This traffic per payment unit $P_m$ is also called the monopoly price in a monopoly market. The payoff of the BS is thus
\begin{equation}
\underline{U}(P_m)=(P_m-c)D(P_m).
\end{equation}
The payoff of the APs is thus
\begin{equation}
\underline{V}(P_m)=\sum_{k=1}^K \beta_k{\theta_k v[d_k(P_m)]-P_md_k(P_m)}.
\end{equation}
The social welfare can be derived by (6).

Taking the first derivative of the payoff of AP $\underline{V}(P_m)$ with respect to $d_k$, we have $\theta_k v'(d_k)=P_m >c$. This result shows that this solution leaves a positive \emph{information rent} for all APs when the BS is unobservable of APs' types. Since the BS can only make a positive payoff by setting the unit traffic $P$ greater than the marginal cost $c$, there is potential idle capacity that has not been utilized to offload traffic, and thus, deteriorate the social welfare. In a monopoly market, the BS sets a higher price than marginal cost, which distorts the trade-offs in the economy and moves it away from Pareto efficiency. The social welfare that has been lost is called the \emph{deadweight loss} from the monopoly.

\subsection{Anti Adverse Selection Contract}

Under the information asymmetry, the BS can receive more payoff than that offering the \emph{linear pricing} which only specifies a traffic per payment unit $P$ to all APs. Indeed, the BS can still offer different traffic-payment bundles as in the \emph{perfect discrimination} case. However, the challenge is that, under the information asymmetry the \emph{perfect discrimination} cannot bring compatible incentive for the APs. When a high type AP selects the traffic-payment bundle intended for a low type AP, the high type AP will receives a positive payoff rather than selecting the bundle designed for itself, which brings it a zero payoff. Thus, the high type APs will pretend to be low type APs, and deteriorate the BS's payoff. Thus, the \emph{perfect discrimination} cannot be extended to the information asymmetry case directly .

To obtain an incentive compatible contract, we propose the \emph{anti adverse selection}, in which the BS still offers $K$ types of traffic-payment bundles as \emph{perfect discrimination} in the symmetric information scenario. In order to prevent high type APs from mimicking the low type APs, we need to add the IC constraint into the optimization problem of the \emph{perfect discrimination}. Therefore, the problem can be expressed as follows:
\begin{eqnarray}\label{eq:Adverse}
&&\!\!\!\!\!\!\!\!\!\!\! \max_{(T,q)} \sum_{k=1}^K \beta_k\left(q_k- cT_k \right),
\\
&&\!\!\!\!\!\!\!\!\!\!\!s.t. \nonumber
\\
&& (a) \quad\!\!\theta_k v(T_k)-q_k \geq 0, \nonumber
\\
&& (b) \quad\!\!\theta_k v(T_k)-q_k \geq \theta_k v(T_l)-q_l,  \nonumber
\\
&& \quad k,l \in \{1, \ldots, K\}, \quad k \neq l. \nonumber
\end{eqnarray}
The optimization problem is constrained by \textit{(a)} and \textit{(b)} which are the IR and IC constraints, respectively. It has been shown that for various parameters, the objective function might be concave or convex, and thus, is not a convex optimization problem \cite{Duan.DySPAN2011}. Here, before we derive the solution, we show several conditions that guarantee the feasibility of the \emph{anti adverse selection} contract.

\begin{remark}
Monotonicity: Having the type of APs following the inequality that $\theta_1< \cdots <\theta_k< \cdots <\theta_K$, BS needs to offer $K$ a contract with different bundles $(T_k,q_k)$ for APs of $K$ types. For each bundle, the offloaded traffic, payment, and AP's payoff have the monotonicity that
\begin{align}
q_1<\cdots<&q_k<\cdots<q_K,  \\
T_1<\cdots<&T_k<\cdots<T_K, \\
0 \leq V(1)< \cdots<&V(k)< \cdots <V(K).
\end{align}
\end{remark}

\begin{proof}
To prove this remark, we first prove that $T_k>T_l$ exists if and only if when $\theta_k>\theta_l$. From the IC constraint in Section III.C, we have
\begin{align}
\theta_k v(T_k)-q_k&> \theta_k v(T_l)-q_l, \\
\theta_l v(T_l) -q_l&> \theta_l v(T_k) -q_k,
\end{align}
with $k,l \in \{1, \ldots, K\}, k \neq l$. We add the two inequalities above together and have:
\begin{align}
\theta_k v(T_k) + \theta_l v(T_l) > \theta_k v(T_l) + \theta_l v(T_k).
\end{align}

If we factor the inequality by $v(T)$, we have
\begin{align}
\theta_k v(T_k) - \theta_l v(T_k) &> \theta_l v(T_k) - \theta_l v(T_l), \\ \nonumber
v(T_k)(\theta_k -\theta_l) &> v(T_l)(\theta_k -\theta_l).
\end{align}
As $\theta_k>\theta_l$, we must have $\theta_k -\theta_l>0$, thus, $v(T_k)>v(T_l)$. From the definition of $v(T)$, we know that $v$ is a strictly increasing function of $T$. As $v(T_k)>v(T_l)$ holds, we must have $T_k>T_l$.

If we factor the inequality by $\theta$, we have
\begin{align}
\theta_k v(T_k) - \theta_k v(T_l) &> \theta_l v(T_k) - \theta_l v(T_l), \\ \nonumber
\theta_k (v(T_k) - v(T_l)) &> \theta_l (v(T_k) - v(T_l)).
\end{align}
As $T_k>T_l$ and $v(T)$ is strictly increasing with $T$, we must have $v(T_k)>v(T_l)$ and $v(T_k) - v(T_l)>0$. Thus, by dividing both sides of the inequality, we get $\theta_k > \theta_l$. As a result, we have proved that $\theta_k>\theta_l$ if and only if $T_k>T_l$.

By now, we have proved the sufficiency and necessity that for any feasible contract $(T,q)$, $T_k>T_l$ if and only if $\theta_k>\theta_l$.
Given our assumption in Definition \ref{Def1} that $\theta_1< \cdots <\theta_k< \cdots <\theta_K$, we have
\begin{equation}
T_1< \cdots <T_k< \cdots <T_K.
\end{equation}
As a strictly increasing function of $T$, the offloading traffic $q$ satisfies the following condition intuitively:
\begin{equation}
q_1<\cdots<q_k<\cdots <q_K.
\end{equation}

When $\theta_k > \theta_l$, we also have
\begin{align}
V(k)=\theta_k v(T_k) -q_k&> \theta_k v(T_l) -q_l \quad  (IC) ,    \\ \nonumber
                           &> \theta_l v(T_l) -q_l=V(l), \\ \nonumber
                           &> \theta_1 v(T_1) -q_1=V(1) \geq 0.
\end{align}
We see that $V(k)>V(l)$ when $\theta_k>\theta_l$. As $\theta_1< \cdots <\theta_k< \cdots <\theta_K$, then
\begin{equation}
0 \leq V(1)< \cdots<V(k)< \cdots <V(K).
\end{equation}
\end{proof}

Monotonicity implies that the APs of higher type, i.e., with the larger idle capacity, receive more payment, together with higher payoff than those of the APs whose types are lower. If a high type AP selects the contract designed for a low type AP, even though a smaller amount of offloaded traffic is required from the BS, the less payment received will deteriorate AP's payoff. Moreover, if a lower type AP selects a traffic-payment bundle intended for a high type AP, it needs to vacate part of the capacity intended for its own traffic which may cause worse cost which will surpass the payment received.

The monotonicity implies that the contract brings compatible incentives for the APs as high capable APs receive more payoff than low capable ones. In addition, it shows that the contract is \emph{self revealing} as the AP can receive the maximum payoff if and only if it selects the contract that best fits into its type. Thus, the problem of high type APs mimicking low type APs is solved. From \cite{Bolton.2004} we see that the \emph{anti adverse selection} specifies a nonlinear traffic for different type APs instead of a linear traffic as in the \emph{linear pricing}, and it brings higher payoffs to the BS and APs, and larger social welfare than the \emph{linear pricing}.

From (\ref{eq:Adverse}) we see that, this problem is not a convex optimization problem. There are $K$ IR constraints and $K(K - 1)$ IC constraints in total. According to \cite{Bolton.2004}, we know that all IR constraints can be removed except
\begin{align}
\theta_1 v(T_1)-q_1 \geq 0.
\end{align}
For the IC constraints, we only need to keep $K-1$ of the $K(K - 1)$ constraints, which are
\begin{align}
\theta_k v(T_k)-q_k \geq \theta_k v(T_{k-1})-q_{k-1}.
\end{align}
With the reduced constraints, we can reformulate the BS's optimization problem as
\begin{eqnarray}\label{eq:ReAdverse}
&&\!\!\!\!\!\!\!\!\!\!\! \max_{(T,q)} \sum_{k=1}^K \beta_k\left(q_k- cT_k \right),
\\
&&\!\!\!\!\!\!\!\!\!\!\!s.t. \nonumber
\\
&& (a) \quad\!\!\theta_1 v(T_1)-q_1 \geq 0, \nonumber
\\
&& (b) \quad\!\!\theta_k v(T_k)-q_k \geq \theta_k v(T_{k-1})-q_{k-1},  \nonumber
\\
&& \quad k,l \in \{1, \ldots, K\}, \quad k \neq l. \nonumber
\end{eqnarray}

We can solve this optimization problem by using Lagrangian multiplier method which is explained in the following steps. First, based on the optimization problem, we have the Largrangian as
\begin{align}
\mathcal{L}&=\sum_{i=1}^K \{\beta_i\left(q_k- cT_k \right) + \mu_i [\theta_k v(T_k)- \theta_k v(T_{k-1}) \nonumber \\
& \quad \quad \quad \quad \quad \quad +q_{k-1}-q_k ] \}+ \nu \theta_1 v(T_1)-q_1.
\end{align}
To find the optimal contract $(T_k,q_k)$, take the partial derivatives regarding $q_k$ and $T_k$, then set the value equal to zero. For $k=K$ we have
\begin{align}
\frac{\partial \mathcal{L}}{\partial T_k} &= \mu_k \theta_k v'(T_k)= \beta_k, \\
\frac{\partial \mathcal{L}}{\partial q_k} &= 0   \quad \Leftrightarrow \beta_k = \mu_k.
\end{align}
For $k\in \{1,\ldots, K-1\}$, we have
\begin{align}
\frac{\partial \mathcal{L}}{\partial T_k} &= \mu_k \theta_k v'(T_k)- \mu_{k+1} \theta_{k+1} v'(T_k)= \beta_k, \\
\frac{\partial \mathcal{L}}{\partial q_k} &= \beta_k - \mu_k +\mu_{k+1}=0.
\end{align}

With the $2K$ equations, we can solve $\mu_k$ and $T_k$, $k\in \{1,\ldots, K\}$, by backward induction. As $\beta_k$, $\theta_k$, and valuation function $v$ are known, we can first get $T_K$ from (38) and (39). $T_k$ for $k\in \{1,\ldots, K-1\}$ are then obtainable by (40) and (41) by solving
\begin{align}
v'(T_k)=\frac{\theta_k}{\mu_k \theta_k - \mu_{k+1} \theta_{k+1}}.
\end{align}
By now, we have the payment $T_k$ available for the optimal contract. The required traffic can be found by using the IC and IR constraints. First, with constraint $(a)$, we have
\begin{align}
q_1 = \theta_1 v'(T_1).
\end{align}
Then, by using constraint $(b)$, we have
\begin{align}
q_k=\theta_i v'(T_k)- \theta_{k} v'(T_{k-1})+q_{k-1},
\end{align}
where $k\in\{2,\dots, K\}$.

The optimal contract solved by this optimization problem gives all APs from type $2$ to $K$ positive payoffs, and only the lowest type of AP will have a zero payoff. The social welfare achieved by the highest type of payoff is the same as as that from the ideal case when the type is known by the BS, which is an efficient transaction. However, the social welfare from the other types is lower than that from the ideal case, and thus yielding an inefficient transaction.
\section{Simulation Results and Analysis}\label{sec:Simulations}

In this section, we will first show the feasibility of the contracts solved from the three mechanisms. Then, we will compare the payoffs of the BS and APs, together with the social welfare of the three contracts. Referring to previous works in contract theory such as \cite{Gao.JSAC2011} and \cite{Duan.DySPAN2011}, we assume $K=20$ and give the simulation with $20$ types of APs. For simplicity, we consider a uniform distribution of AP types, i.e., $\beta_k=1/K$. We set the unit payment cost of the BS $c=0.01$.

\subsection{Contract Feasibility}

\subsubsection{Monotonicity}
\begin{figure}
 \begin{subfigure}[b]{0.45\textwidth}
 \includegraphics[width=\columnwidth,height=0.8\textwidth]{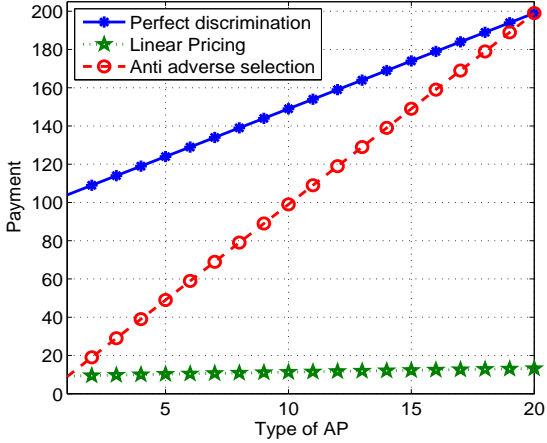}
 \caption{Different type AP's request of payment.}
 \label{fig:payment}
 \end{subfigure}
 \begin{subfigure}[b]{0.45\textwidth}
 \includegraphics[width=\columnwidth,height=0.8\textwidth]{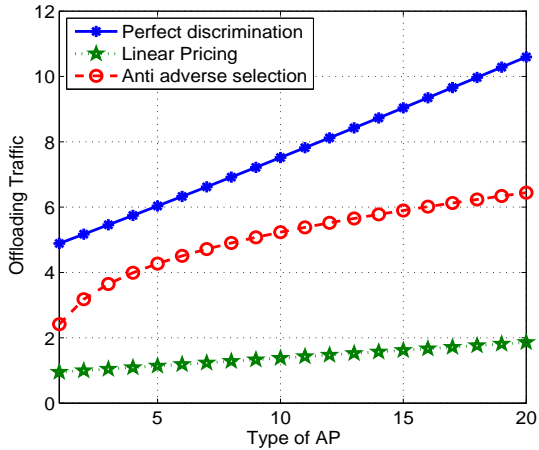}
 \caption{BS's requirement of offloading traffic for different type APs.}
 \label{fig:traffic}
 \end{subfigure}
\caption{Incentive compatibility of the three contracts.}
\label{fig:contract}
\vspace{-0.2cm}
\end{figure}

In Fig.\ref{fig:contract}, we show the payment and required traffic in the contracts given by the three proposed mechanisms are \emph{incentive compatible}, which ensures that the payment and offloading traffic are proportional with the APs' types.

In Fig. \ref{fig:payment}, we plot the payment required from the APs to the BS of the three contracts. We see that in all three contracts, the APs of high type always require more payment than that of the low type APs, i.e., the payment required by the AP is a strictly increasing function of the AP's type. The highest payment shown in Fig. \ref{fig:payment} also proves our conclusion that the highest type AP in \emph{anti adverse selection} achieves the optimal efficiency as in the \emph{perfect discrimination} contract.

\begin{figure}[t]
  \begin{center}
    \includegraphics[width=\columnwidth,height=0.35\textwidth]{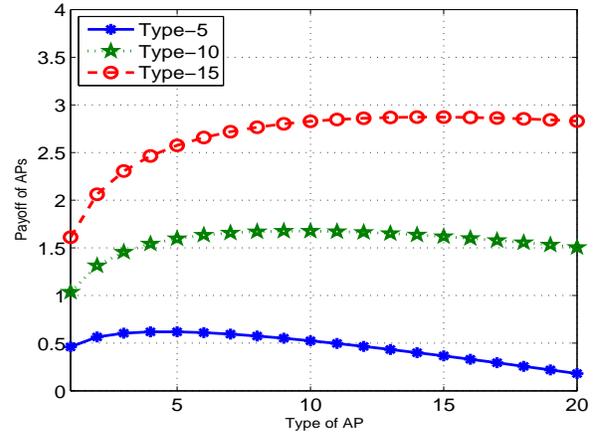}
    \caption{\label{fig:select}AP's payoff with different type traffic-payment bundles.}
  \end{center}
\vspace{-0.2cm}
\end{figure}
Fig. \ref{fig:traffic} gives the amount of offloaded traffic required from the BS to different type APs. It is clear that the amount of offloaded traffic is also a strictly increasing function of the AP type, i.e., a higher type AP (with large idle capacity) is required to offload more cellular than that of a low type AP (with small idle capacity) such that the workload is in accordance with the payment they received. Furthermore, we see that the required amount of traffic in \emph{perfect discrimination} and \emph{linear pricing} are linear functions of type, and it is a concave function of type in \emph{anti adverse selection} which is consistent with our conclusion in Section IV.C that the offloading traffic $q$ is a nonlinear curve. Among the three contracts, the \emph{perfect discrimination} contract requires the largest amount of offloaded traffic from the AP, followed by the \emph{anti adverse selection} under information asymmetry. The smallest amount of traffic is required under \emph{linear pricing}.

We have seen similar results from \cite{Gao2014} that the price in a perfect competition market and the monopoly price serves as the upper and lower bounds of market price, respectively. The reason is that the price competition in a competitive market drives the market price up and the monopoly market drives the market price down as the monopolist is selfish and only tries to maximize its own payoff.

\subsubsection{Self Revealability}

In Fig. \ref{fig:select}, we show that the contract solved by \emph{anti adverse selection} is self revealing. We plot the payoffs of \emph{type-5}, \emph{type-10}, and \emph{type-15} APs when selecting all the traffic-payment bundles offered by the BS. In Fig. \ref{fig:select} we see that the payoffs of each AP are concave functions. The maximum points of the three curves are at the point when APs selecting the type of traffic-payment bundle that is the same as its own type. This proves that the contract solved by \emph{anti adverse selection} can automatically reveal the real type of the APs. Thus, by designing a contract in this form, the type of an AP will be automatically revealed to the BS after its selection. In other words, the optimal contract under information asymmetry enables the BS to solve the information asymmetry and retrieve the information related to AP type.

Moreover, Fig. \ref{fig:select} shows that the payoffs of the three types of APs follows the inequality $u_5 <u_{10} < u_{15}$ when they select the same traffic-payment bundle. This corroborates the result shown in the monotonicity condition that the higher the type of the AP, the larger the payoff it can receive when selecting the same contract.

\subsection{System Performance}

\begin{figure}[!t]
  \begin{center}
    \includegraphics[width=\columnwidth,height=0.35\textwidth]{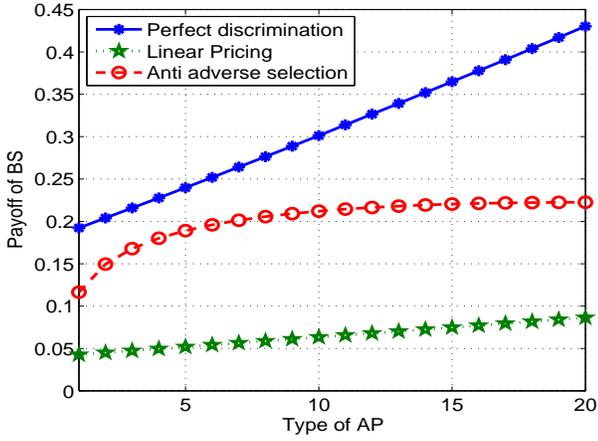}
    \caption{\label{fig:BS}BS's payoff when contracting with different type APs.}
  \end{center}
 \vspace{-0.5cm}
\end{figure}

First, in Fig. \ref{fig:BS} we show the payoff of the BS under the three contracts. The BS receives the maximum payoff when there is no information asymmetry in the \emph{perfect discrimination}, since the BS has full knowledge of AP types. Nonetheless, we can see that the proposed solution with information asymmetry by the \emph{anti adverse selection} yields a payoff for the BS that outperforms the \emph{linear pricing} which receives the lowest payoff. Furthermore, the BS always receive higher payoff when trading with high type APs than that with low type APs. This result also in accordance with our conclusion that the BS prefers to trade with high type APs than with low type APs.

Here, we note that, even though the \emph{anti adverse selection} contract under information asymmetry can force the APs to reveal their types, the exact value of the AP type is still unavailable to the BS. Thus, the BS can only achieve a near optimal payoff under information asymmetry, which is always upper bounded by the \emph{perfect discrimination}. The \emph{linear pricing} contract does not impose any restriction on the APs' choice of contract and less information is retrieved, which impedes the BS from obtaining more payoff.

\begin{figure}[!t]
  \begin{center}
    \includegraphics[width=\columnwidth,height=0.35\textwidth]{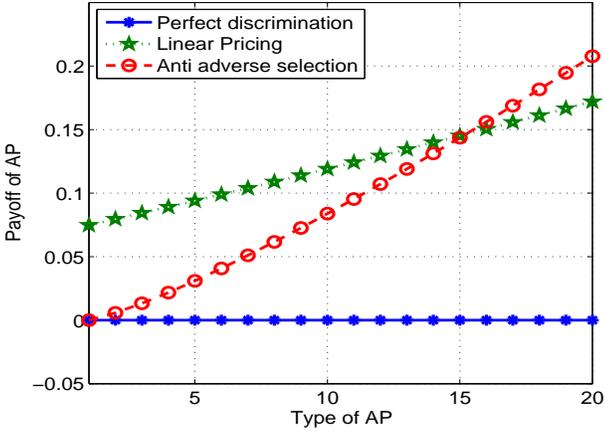}
    \caption{\label{fig:AP}Different type APs' payoff.}
  \end{center}
\vspace{-0.5cm}
\end{figure}

In Fig. \ref{fig:AP}, we plot the payoff of different type APs. In the \emph{linear contract} and \emph{anti adverse selection}, the payoff of APs is increasing with the type. By contrast, the payoff of APs remains zero in the \emph{perfect discrimination}. This result proves our conclusion in Section IV.A that when the BS is available of the AP's type, it will not leave any \emph{information rent} for the AP. The \emph{anti adverse selection} contract proves the monotonicity of the contract that the higher the type of AP, the larger the payoff it can receive under information asymmetry. Furthermore, we see that all type APs in \emph{anti adverse selection} enjoys a positive payoff which is the representation of \emph{information rent} except for the lowest type $\theta_1$ AP.

Overall, we see that \emph{linear pricing} gives the APs the highest payoff, followed by the \emph{anti adverse selection} under information asymmetry, then the ideal \emph{perfect discrimination} with no information asymmetry. However, for some of the high type APs can obtain higher payoff from the optimal contract under information asymmetry than the \emph{linear pricing}.

\begin{figure}[!t]
  \begin{center}
    \includegraphics[width=\columnwidth,height=0.35\textwidth]{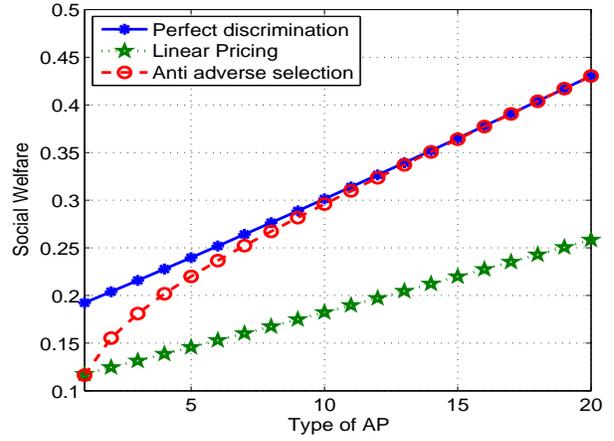}
    \caption{\label{fig:social}Social welfare when BS contracting with different type APs.}
  \end{center}
\vspace{-0.5cm}
\end{figure}

In Fig. \ref{fig:social}, we show the social welfare of the three contracts when the BS trading with different type APs. The social welfare is also an increasing function of the AP's type. The \emph{linear pricing} loses more social welfare than the \emph{anti adverse selection contract} when comparing with the \emph{perfect discrimination}. The \emph{anti adverse selection} brings a low utilization of APs' idle capacity compare to the \emph{perfect discrimination}, except for the highest type $\theta_K$ AP. The efficiency of idle capacity utilization is only kept for the top type APs. We can name this situation as the idle capacity utilization is \emph{distorted}. The \emph{linear pricing} contract gives the lowest social welfare (i.e., trading efficiency) since no in formation retrieving strategy has been apply.

The \emph{distortion} of lower type APs' offloading traffic in the \emph{anti adverse selection} is the result of the BS's attempt to reduce the \emph{information rent} of the high type APs. As we have mentioned at the beginning of of Section IV.C, when there is an information asymmetry, the high type APs have the incentive to mimic low type APs to receive more payoff. In order to reduce high type APs' incentives in doing this, the BS lowers the required amount of traffic of low type APs. Since high type APs prefer to offload traffic for the BS, this action stops the high type APs from pretending to be a low type AP, meanwhile, leading to an underuntilization of low type AP's idle capacity.

\begin{figure*}
 \begin{subfigure}[b]{0.32\textwidth}
 \centering
 \includegraphics[width=\columnwidth,height=0.75\textwidth]{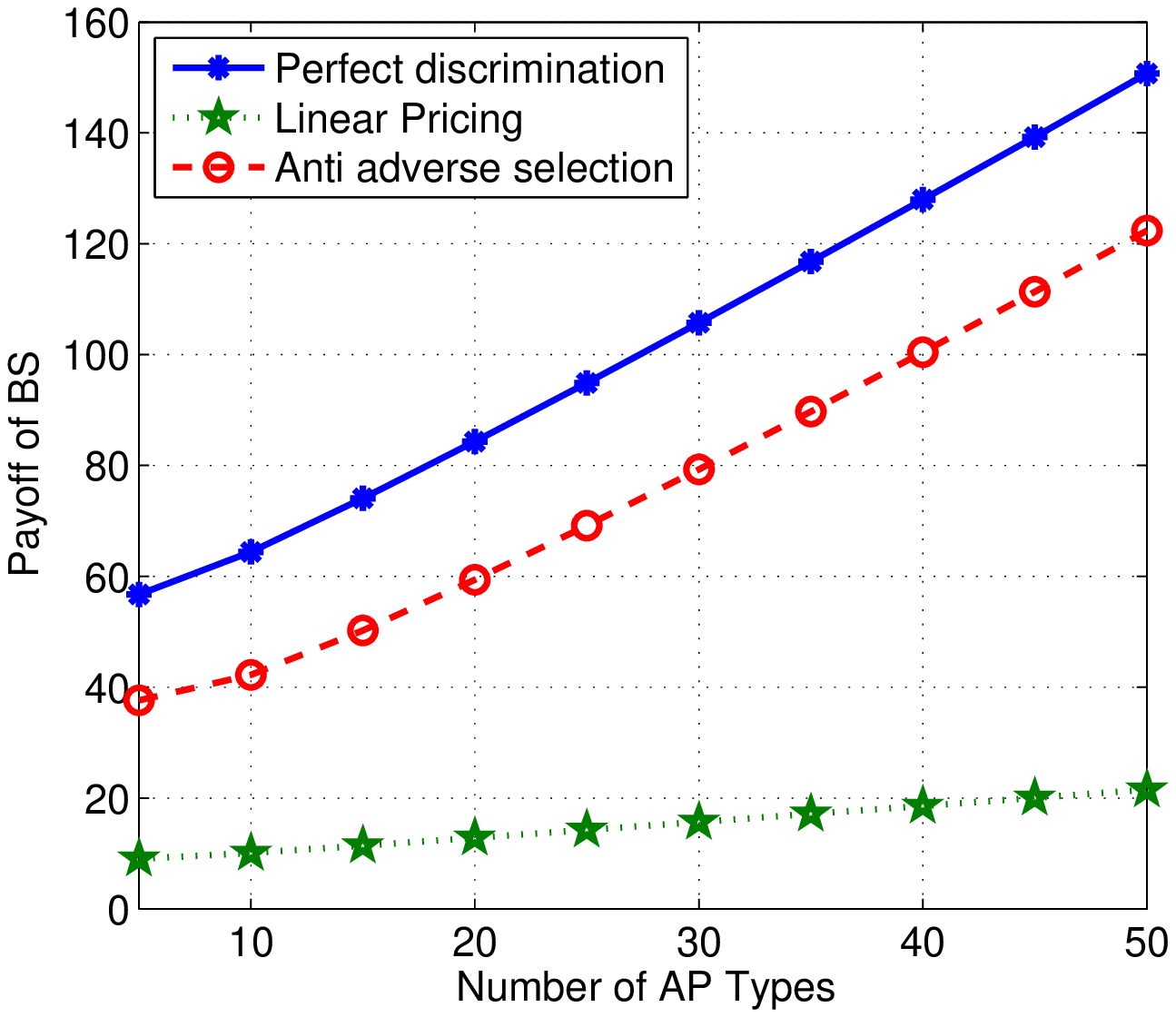}
 \caption{Payoff of BS.}
 \label{fig:numBS}
 \end{subfigure}
 \begin{subfigure}[b]{0.32\textwidth}
 \centering
 \includegraphics[width=\columnwidth,height=0.75\textwidth]{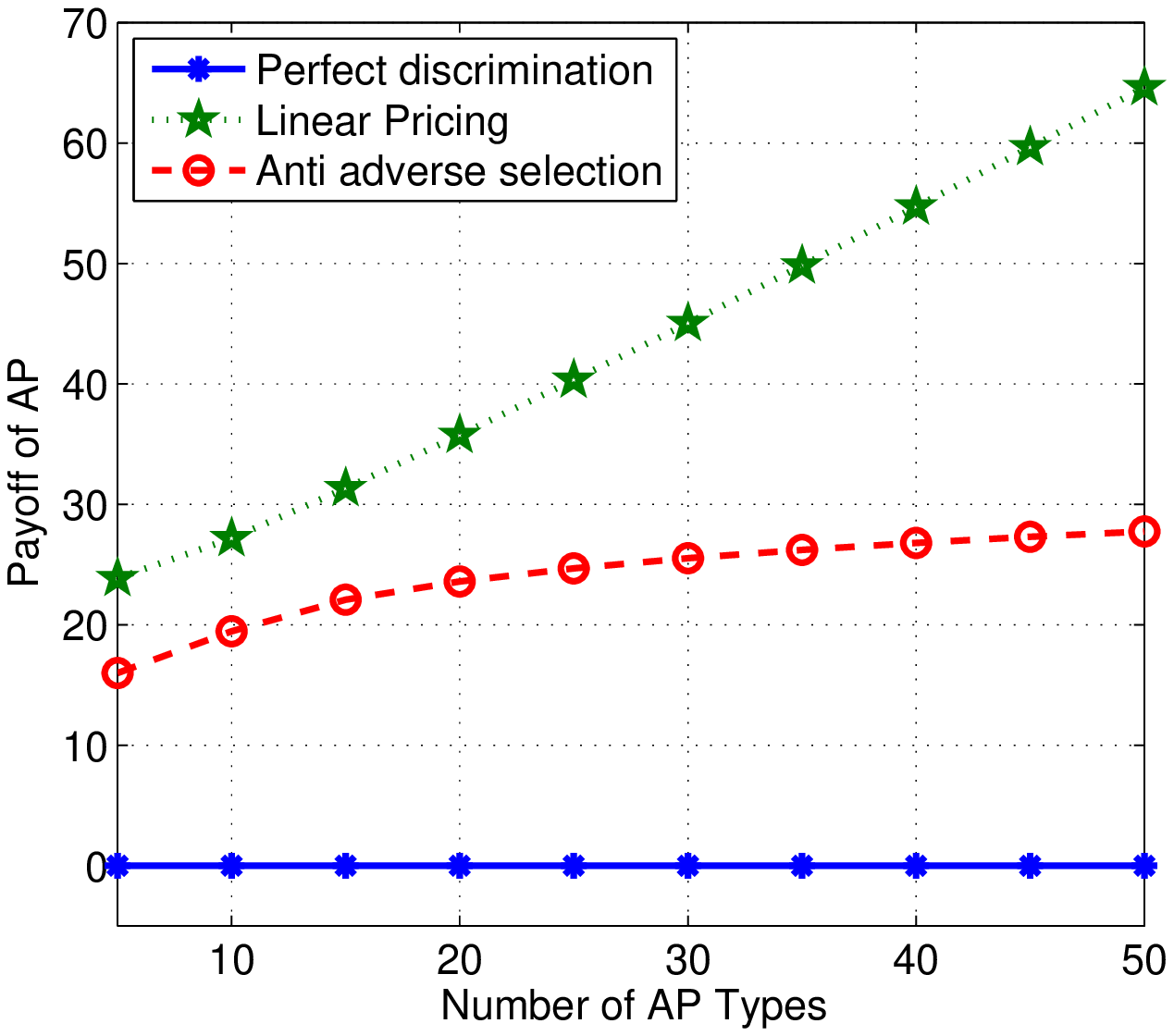}
 \caption{Payoff of AP.}
 \label{fig:numAP}
 \end{subfigure}
  \begin{subfigure}[b]{0.32\textwidth}
 \centering
 \includegraphics[width=\columnwidth,height=0.75\textwidth]{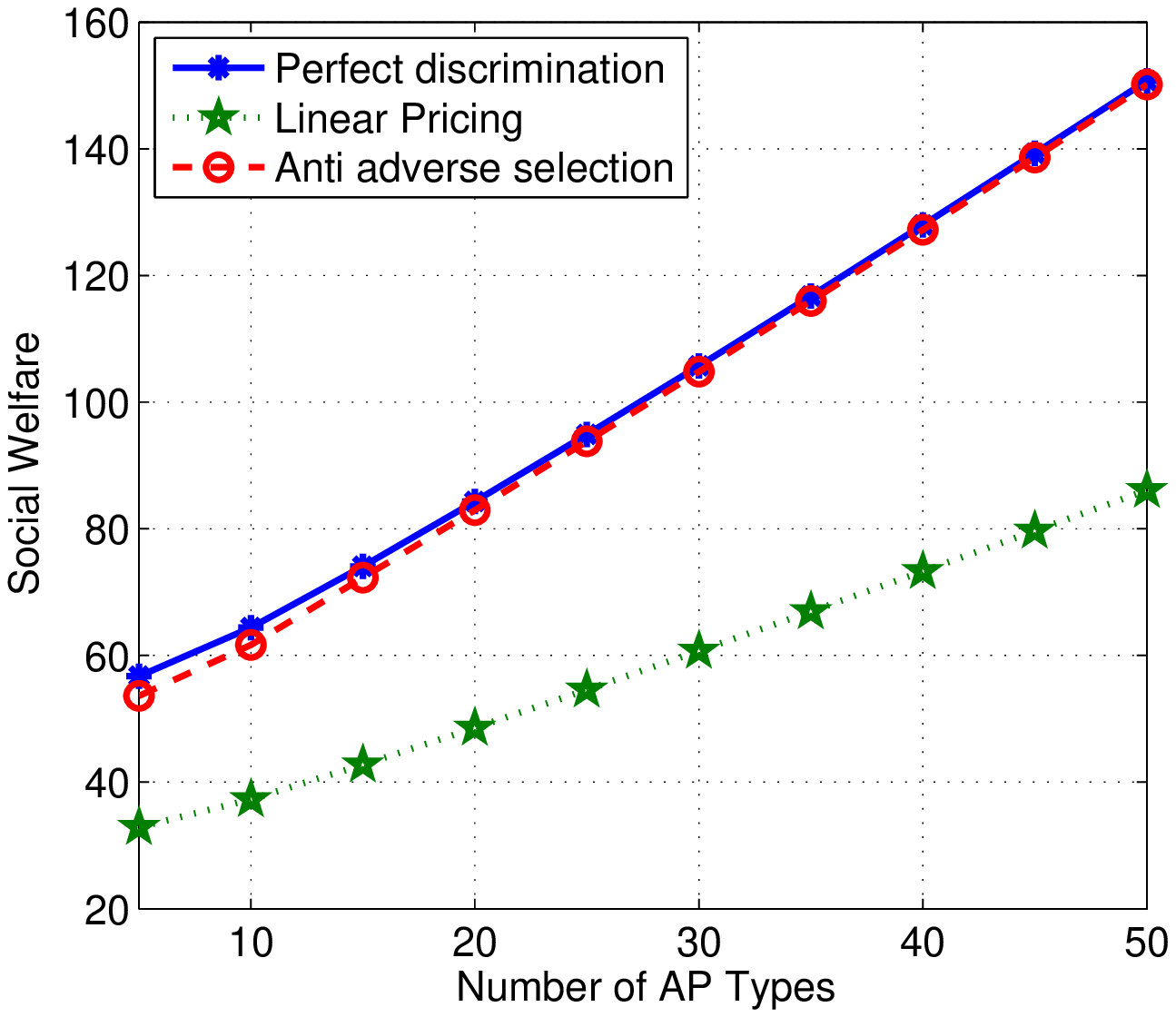}
 \caption{Social Welfare.}
 \label{fig:numSW}
 \end{subfigure}
\centering
\caption{The system performance when the number of AP types varies.}
\label{fig:num}
\vspace{-0.3cm}
\end{figure*}
In Fig. \ref{fig:num}, we study the system performance when the number of AP types increases, while the other parameters are fixed. An increase in the number of types will inherently yield an increase in the total number of APs pairs. Fig. \ref{fig:numBS} shows similar properties as Fig. \ref{fig:BS}, since the BS can extract all payoff from the APs due to the symmetric information in \emph{perfect discrimination}, while no \emph{information rent} is retried by \emph{linear pricing}. Fig. \ref{fig:numAP} also proves the conclusions that we have made in Fig. \ref{fig:AP} that, when the BS cannot retrieve \emph{information rent} from the APs in \emph{linear pricing}, the APs receive the highest payoff, and the APs cannot gain any payoff when the BS has full information in \emph{perfect discrimination}.

In Fig. \ref{fig:numSW}, we show the social welfare of the three contracts. We see that the social welfare is an increasing function with the number of AP types. From Fig. \ref{fig:numSW} we also see that without information asymmetry, the \emph{perfect discrimination} achieves the first best social welfare as we have stated that it brings the Perato optimal for the system. Followed by the \emph{anti adverse selection} and \emph{linear pricing} under the information asymmetry. The larger the number of APs, the closer the \emph{anti adverse selection} contract approximates the \emph{perfect discrimination}.

\section{Conclusions}\label{sec:Conclusion}

In this paper, we have presented a contract-theoretic model for addressing the problem of incentivizing APs to offload traffic for the BSs in an SDN-at-edge. We have proposed a self-revealing contract: \emph{anti adverse selection} to overcome the information asymmetry that the BS is unaware of the AP's idle capacity. To evaluate the efficiency of our proposed contract, we have introduced and analysed the \emph{perfect discrimination} and \emph{linear pricing} as comparisons. Simulation results have shown that our proposed approaches can effectively incentivize APs to participate in the traffic offloading process. In addition, the \emph{anti adverse selection} can achieve the second best outcome compare to the ideal case \emph{perfect discrimination}, while gives an optimal social welfare compare with the \emph{linear pricing} under information asymmetry.

\bibliographystyle{IEEEtran}
\bibliography{./JSPS}

\begin{thebibliography}{10}
\providecommand{\url}[1]{#1}
\csname url@samestyle\endcsname
\providecommand{\newblock}{\relax}
\providecommand{\bibinfo}[2]{#2}
\providecommand{\BIBentrySTDinterwordspacing}{\spaceskip=0pt\relax}
\providecommand{\BIBentryALTinterwordstretchfactor}{4}
\providecommand{\BIBentryALTinterwordspacing}{\spaceskip=\fontdimen2\font plus
\BIBentryALTinterwordstretchfactor\fontdimen3\font minus
  \fontdimen4\font\relax}
\providecommand{\BIBforeignlanguage}[2]{{%
\expandafter\ifx\csname l@#1\endcsname\relax
\typeout{** WARNING: IEEEtran.bst: No hyphenation pattern has been}%
\typeout{** loaded for the language `#1'. Using the pattern for}%
\typeout{** the default language instead.}%
\else
\language=\csname l@#1\endcsname
\fi
#2}}
\providecommand{\BIBdecl}{\relax}
\BIBdecl

\bibitem{Sesia2009}
S.~Sesia, I.~Toufik, and M.~Baker, \emph{{LTE}: the UMTS long term
  evolution}.\hskip 1em plus 0.5em minus 0.4em\relax New York: John Wiley \&
  Sons, 2009.

\bibitem{Cisco2014}
Cisco, ``Cisco visual networking index: Global mobile data traffic forecast
  update, 2013-2018,'' Tech. Rep., Feb. 2014.

\bibitem{Manzalini2013}
A.~Manzalini and R.~Saracco, ``Software networks at the edge: A shift of
  paradigm,'' in \emph{Future Networks and Services (SDN4FNS), IEEE SDN for},
  Berlin, Germany, Nov. 2013.

\bibitem{Ali-Ahmad2013}
H.~Ali-Ahmad, C.~Cicconetti, A.~de~la Oliva, V.~Mancuso, M.~Reddy~Sama,
  P.~Seite, and S.~Shanmugalingam, ``An {SDN-Based} network architecture for
  extremely dense wireless networks,'' in \emph{Future Networks and Services
  (SDN4FNS), IEEE SDN for}, Berlin, Germany, Nov. 2013.

\bibitem{Gudipati2013}
A.~Gudipati, D.~Perry, L.~E. Li, and S.~Katti, ``{SoftRAN}: Software defined
  radio access network,'' in \emph{HotSDN}, Hong Kong, China, Aug. 2013.

\bibitem{Galis2013}
A.~Galis, S.~Clayman, L.~Mamatas, J.~Rubio~Loyola, A.~Manzalini, S.~Kuklinski,
  J.~Serrat, and T.~Zahariadis, ``Softwarization of future networks and
  services -programmable enabled networks as next generation software defined
  networks,'' in \emph{Future Networks and Services (SDN4FNS), IEEE SDN for},
  Berlin, Germany, Nov. 2013.

\bibitem{ONF2013}
\BIBentryALTinterwordspacing
O.~S. Brief, ``{OpenFlow}-enabled mobile and wireless networks,'' white paper,
  2013. [Online]. Available:
  \url{https://www.opennetworking.org/images/stories/downloads/sdn-resources/solution-briefs/sb-wireless-mobile.pdf}
\BIBentrySTDinterwordspacing

\bibitem{Werin.1992}
L.~Werin and H.~Wijkander, \emph{Contract Economics}.\hskip 1em plus 0.5em
  minus 0.4em\relax Oxford, UK: Blackwell Publishers, 1992.

\bibitem{Zhang.Glob2013}
Y.~Zhang, L.~Song, W.~Saad, Z.~Dawy, and Z.~Han, ``Exploring social ties for
  enhanced device-to-device communications in wireless networks,'' in
  \emph{IEEE Globe Communication Conference (Globecom)}, Atlanta, GA, Dec.
  2013.

\bibitem{Camps2013}
D.~Camps-Mur, A.~Garcia-Saavedra, and P.~Serrano, ``Device-to-device
  communications with {Wi-Fi} direct: overview and experimentation,''
  \emph{Wireless Communications, IEEE}, vol.~20, no.~3, pp. 96--104, Jun. 2013.

\bibitem{Quek.2013}
T.~Q.~S. Quek, G.~de~la Roche, I.~Guvenc, and M.~Kountouris, \emph{Small Cell
  Networks: Deployment, {PHY} Techniques,and Resource Management}.\hskip 1em
  plus 0.5em minus 0.4em\relax UK: Cambridge University Press, 2013.

\bibitem{Oh2011}
E.~Oh, B.~Krishnamachari, X.~Liu, and Z.~Niu, ``Toward dynamic energy-efficient
  operation of cellular network infrastructure,'' \emph{IEEE Commun. Mag.},
  vol.~49, no.~6, pp. 56--61, Jun. 2011.

\bibitem{Valdivieso2013}
A.~Valdivieso~Caraguay, L.~Barona~Lopez, and L.~Garcia~Villalba, ``Evolution
  and challenges of software defined networking,'' in \emph{Future Networks and
  Services (SDN4FNS), IEEE SDN for}, Berlin, Germany, Nov 2013.

\bibitem{Mueller2013}
J.~Mueller, A.~Wierz, and T.~Magedanz, ``Scalable on-demand network management
  module for software defined telecommunication networks,'' in \emph{Future
  Networks and Services (SDN4FNS), IEEE SDN for}, Berlin, Germany, Nov. 2013.

\bibitem{Gao2014}
L.~Gao, G.~Iosifidis, J.~Huang, and L.~Tassiulas, ``Hybrid data pricing for
  network-assisted user-provided connectivity,'' in \emph{INFOCOM, Proceedings
  IEEE}, Toronto, Canada, Apr. 2014.

\bibitem{Paris2013}
S.~Paris, F.~Martignon, I.~Filippini, and L.~Chen, ``A bandwidth trading
  marketplace for mobile data offloading,'' in \emph{INFOCOM, Proceedings
  IEEE}, Turin, Italy, Apr. 2013.

\bibitem{Gao2013}
L.~Gao, G.~Iosifidis, J.~Huang, and L.~Tassiulas, ``Economics of mobile data
  offloading,'' in \emph{INFOCOM, Proceedings IEEE}, Turin, Italy, Apr. 2013.

\bibitem{Iosifidis2014}
G.~Iosifidis, L.~Gao, J.~Huang, and L.~Tassiulas, ``Enabling crowd-sourced
  mobile internet access,'' in \emph{INFOCOM, Proceedings IEEE}, Toronto,
  Canada, Apr. 2014.

\bibitem{KnapperCEC11}
R.~Knapper, B.~Blau, T.~Conte, A.~Sailer, A.~Kochut, and A.~Mohindra,
  ``Efficient contracting in cloud service markets with asymmetric information
  - a screening approach,'' in \emph{IEEE 13th Conference on Commerce and
  Enterprise Computing (CEC)}, Luxembourg, Sep. 2011.

\bibitem{Duan.INFOCOM2012}
L.~Duan, T.~Kubo, K.~Sugiyama, J.~Huang, T.~Hasegawa, and J.~Walrand,
  ``Incentive mechanisms for smartphone collaboration in data acquisition and
  distributed computing,'' in \emph{The 31st IEEE International Conference on
  Computer Communications (INFOCOM)}, Orlando, FL, Mar. 2012.

\bibitem{Gao.JSAC2011}
L.~Gao, X.~Wang, Y.~Xu, and Q.~Zhang, ``Spectrum trading in cognitive radio
  networks: A contract-theoretic modeling approach,'' \emph{IEEE Journal on
  Selected Areas in Communications (JSAC)}, vol.~29, no.~4, pp. 843--855, Apr.
  2011.

\bibitem{Gao.JSAC2013}
L.~Gao, J.~Huang, Y.~Chen, and B.~Shou, ``An integrated contract and auction
  design for secondary spectrum trading,'' \emph{IEEE Journal on Selected Areas
  in Communications (JSAC)}, vol.~31, no.~3, pp. 581--592, Mar. 2013.

\bibitem{Duan.DySPAN2011}
------, ``Contract-based cooperative spectrum sharing,'' \emph{IEEE
  Transactions on Mobile Computing}, vol.~13, no.~1, pp. 174--187, Jan. 2014.

\bibitem{Bolton.2004}
P.~Bolton and M.~Dewatripont, \emph{Contract Theory}.\hskip 1em plus 0.5em
  minus 0.4em\relax Cambridge, MA: The MIT Press, 2004.

\end{thebibliography}

\end{document}